\documentclass[a4paper,10.5pt]{article}
\usepackage{theorem}
\usepackage{latexsym,amssymb,amsfonts,amsmath,cite}
\usepackage[dvips]{graphicx}
\newtheorem{theorem}{Theorem}
\newtheorem{lemma}{Lemma}
\newtheorem{corollary}[theorem]{Corollary}

\newcommand{\bs}[1]{\boldsymbol{#1}}
\newcommand{\weight}{\boldsymbol{w}^{(\varphi_0)}}

\renewcommand{\labelenumi}{(\theenumi)}

\title{{\Large {\bf On the entropy of decoherence matrix \\ for quantum walks 
}
}}

\author{ 
{\small 
Norio Konno,$^{1}$ 
\footnote{konno@ynu.ac.jp 
}\quad 
Etsuo Segawa,$^{2}$ 
\footnote{segawa@t.stat.u-tokyo.ac.jp
}\quad  
}\\ 
{\scriptsize $^{1}$ 
Department of Applied Mathematics, Faculty of Engineering, Yokohama National University
}\\
{\scriptsize Hodogaya, Yokohama 240-8501, Japan
} \\
{\scriptsize $^2$ 
Department of Mathematical Informatics, The University of Tokyo,
}\\
{\scriptsize Bunkyo, Tokyo, 113-8656, Japan
} \\
} 

\date{\empty }
\begin{document}
\maketitle

\par\noindent
\begin{small}
\par\noindent
{\bf Abstract}. The decoherence matrix studied by Gudder and Sorkin (2011) 
can be considered as a map from the set of all the pairs of $n$-length paths to complex numbers, 
which is induced by the discrete-time quantum walk. 
The decoherence matrix is one of the decoherence functionals which present their historical quantum measure theory. 
In this paper, we compute the von Neumann entropy of the decoherence matrix. 
To do so, we use the result that the eigensystem of the decoherence matrix can be expressed by a corresponding correlated random walk. 


\end{small}

\setcounter{equation}{0}
\section{Introduction}
The quantum measure ($q$-measure) spaces were introduced by Sorkin~\cite{Sorkin} for an approach to quantum mechanics 
and applications to quantum gravity and cosmology. 
Gudder blushed up the construction of the $q$-measure in Ref.\cite{Gudder2}. 
As its consequence, a treatment of $q$-measure infinite spaces has been ensured in his general definition. 
Recently, Gudder and Sorkin \cite{GS} presented a decoherence matrix on $n$-path space $\Omega_n=\{-1,1\}^{\otimes n}$ 
to give a $q$-measure on the event of $\Omega_n$. 
The decoherence matrix is induced by trajectories of a particle of the discrete-time quantum walk (QW) 
which was originated by the first author Gudder~\cite{Gudder1} (1988). 
There are some other opinions of the priority of the QW, for example, Refs.\cite{Aharonov,Meyer}. 
Either way, the QW has been intensively investigated from various view points since around 2000, for example, 
quantum search algorithms~\cite{ABNVW,Ambainis,Szegedy,Magniez}, fundamental physics~\cite{Strauch,CKSS,KMW}, 
limit theorems for its statistical behaviors~\cite{KonnoBook,KLS}, the Anderson localization~\cite{Joye,Hanover,OK}, spectral analysis~\cite{CGMV,KS}, 
its experimental implementations~\cite{Karski_ETAL}, and photonic synthesis~\cite{MohseniETAL}. 
To emphasize an aspect of the $q$-measure theory, 
Gudder and Sorkin~\cite{GS} have mainly considered the decoherence matrix induced by a simple model, that is, two-site QW 
with a special quantum coin and initial coin state. 

In this paper, we generalize their definition of the decoherence matrix 
so that QWs on $\mathbb{Z}$ with general quantum coins and initial coin states can be also treated. 
More precisely, for a given subset $A\subseteq \Omega_n\times \Omega_n$, 
our decoherence matrix $D_A: \Omega_n\times \Omega_n \to \mathbb{C}$ is restricted to $A$ 
in that for any $(\xi,\eta)\notin A$, $D_A(\xi,\eta)=0$. 
We call such $A\subseteq \Omega_n\times \Omega_n$ ``restriction subset". 
We treat the following three kinds of restriction subsets $A_0$, $A_P$ and $A_1$. 
At first, as we will see later, the decoherence matrix restricted to $A_0$ is isomorphic to the decoherence matrix of the two-site QW~\cite{GS} in the end. 
The second subset $A_P$ corresponds to the QW itself on $\mathbb{Z}$, 
that is, $\sum_j\xi_j=\sum_j\eta_j$ for every $(\xi,\eta)\in A_P$. 
In a classical case, there are no correlations between different two paths. 
That is, no correlation effects appear as the diagonal elements of the decoherence matrix. 
To see this, the third subset $A_1$ is the set of every path itself. 
In the view point of the decoherence matrix, we should remark that $A_0 \supseteq A_P \supseteq A_1$. 
In this paper, we compute the von Neumann entropy of the decoherence matrices restricted to the three cases of subsets $A_0$, $A_P$, and $A_1$, respectively. 
Denote the von Neumann entropy by $S_A$ for each case, respectively ($A\in\{A_0,A_P,A_1\}$). 
We find that each eigenvalue of $D_{A}$ ($A\in\{A_0,A_PA_1\}$) is expressed by 
the probability that a particle of the correlated random walk~\cite{Konno_Cor} 
walks along with an $n$-truncated path corresponding to the restriction subset $A$, respectively. 
We find that as the cardinality of the restriction subset is smaller, then the von Neumann entropy becomes larger, 
indeed, $S_{A_0}\sim 1$, $S_{A_P}\sim \log n$, $S_{A_1}\sim n$ for large $n$, since $A_0\supset A_P\supset A_1$.

This paper is organized as follows. 
Section 2 proposes the definition of the decoherence matrix. 
Each von Neumann entropy of the decoherence matrices restricted by $A_0$, $A_P$ and $A_1$, respectively, 
is presented in Sect.3. 
Its proofs are devoted in Sect.4. 
Finally we give the discussion in Sect.5. 

\section{Definition of decoherence matrix for QW}
Let $\Omega_n\equiv \{-1,1\}^{\otimes n}$ be the set of all the $n$-truncated paths. 
Prepare a 2-dimensional unitary matrix called quantum coin as 
\begin{equation}\label{qc}
U=\begin{bmatrix} a & b \\ c & d \end{bmatrix} 
\end{equation}
with $a,b,c,d\in \mathbb{C}$ and $abcd\neq 0$. 
Here $\mathbb{C}$ is the set of complex number. 
We should remark that from the unitarity of $U$, $|a|^2=|d|^2=1-|b|^2=1-|c|^2$ and $d=\Delta \bar{a}$, $c=-\Delta \bar{b}$, 
where $\Delta$ is the determinant of $U$, and $\bar{z}$ is the conjugate of $z\in \mathbb{C}$.
In this paper we denote $\bs{e}_{-1}={}^T[1,0]$, $\bs{e}_1={}^T[0,1]$ corresponding to left and right chiralities, respectively.
For $\bs{\varphi}_0\in \mathbb{C}^2$ with $||\bs{\varphi}_0||=1$ called initial state, 
we define a map $\bs{w}^{(\varphi_0)}: \Omega_n \to \mathbb{C}^2$ such that 
\begin{equation}
\bs{w}^{(\varphi_0)}(\xi)=P_{\xi_n}\cdots P_{\xi_1}\bs{\varphi}_0 
\end{equation}
for all $\xi=(\xi_n,\cdots,\xi_1)\in \Omega_n$ with $\xi_j\in \{-1,1\}$ ($j\in \{1,\dots,n\}$), 
where $P_{\xi_j}=\bs{e}^\dagger_{\xi_j} \bs{e}_{\xi_j} U$. We call $\bs{w}^{(\varphi_0)}(\xi)$ weight of path $\xi$. 
For any $A\subset \Omega_n \times \Omega_n$, we also define a map $D_A: \Omega_n\times \Omega_n \to \mathbb{C}$ such that 
\begin{equation}
D_A(\xi,\eta)=I_{\{(\xi,\eta)\in A\}}(\xi,\eta) \left\langle \bs{w}^{(\varphi_0)}(\xi), \bs{w}^{(\varphi_0)}(\eta) \right\rangle, \;\;\;(\xi,\eta\in \Omega_n), 
\end{equation}
where $I_{\{(\xi,\eta)\in A\}}(\xi,\eta)$ is the indicator function, that is, $I_{\{(\xi,\eta)\in A\}}(\xi,\eta)=1$ $((\xi,\eta)\in A)$, $=0$ $((\xi,\eta)\notin A)$. 
We call $D_A$ decoherence matrix restricted to a set $A\subseteq \Omega_n\times \Omega_n$. \\
\section{Main results}
We introduce an inclusion relation between two subsets of $\Omega_n \times \Omega_n$ in the view point of the decoherence matrix: 
for any $A,B\subset \Omega_n\times\Omega_n$, $A\prec{B}$ means that if $D_B(\xi,\eta)=0$, then $D_A(\xi,\eta)=0$ for $\xi,\eta\in\Omega_n$. 
In particular, if $A\prec{B}$ and $A\succ {B}$, then we denote $A\approx B$. 
In this section, we consider the von Neumann entropy restricted 
by the following three subsets $A_1^{(n)}\prec A_P^{(n)} \prec A_0^{(n)}$ of $\Omega_n\times \Omega_n$ :
\begin{enumerate}
\item
$A_0^{(n)}=\{(\xi,\eta)\in \Omega_n\times\Omega_n: \xi_n=\eta_n \} $ 

We should remark that for any $\xi\in \Omega_n$, there exists $c\in \mathbb{C}$ such that 
\begin{equation}\label{pero}
\bs{w}^{(\varphi_0)}(\xi)=c\bs{e}_{\xi_n}. 
\end{equation}
The last direction of the path appears as the chirality of its weight of the path. 
So the subset $A_0$ is maximal in the following meaning: 
for any $A'\subset \Omega_n\times \Omega_n$ with $A'\supset A_0$, we see $A_0 \approx A'$, 
since $\langle \bs{w}^{(\varphi_0)}(\xi),\bs{w}^{(\varphi_0)}(\eta) \rangle=0$ for any $\xi,\eta\in \Omega_n$ with $\xi_n\neq \eta_n$. 

From now on, we review the original definition of the two-site QW introduced by Gudder and Sorkin \cite{GS}. 
The amplitude that a particle moving between two sites ``$0$" and ``$1$" starting from the site ``$0$" stays the same site is $1/\sqrt{2}$, and 
the amplitude that the particle changes its present site is $i/\sqrt{2}$ for each time step. 
Let $\omega=(\alpha_n,\dots,\alpha_{0})$ and $\omega'=(\alpha_n',\dots,\alpha_{0}')$, $(\alpha_j,\alpha_j'\in \{0,1\})$ with $\alpha_0=\alpha_0'=0$ 
be the two $n$-length trajectories in the two-site walk. The joint amplitude between $\omega$ and $\omega'$ is defined by 
\begin{equation}\label{umetani} 
D(\omega,\omega')=I_{\{\alpha_n=\alpha_n'\}}(\omega,\omega')
	\left(\frac{i^{|\alpha_0-\alpha_1|}}{\sqrt{2}}\cdots \frac{i^{|\alpha_{n-1}-\alpha_{n}|}}{\sqrt{2}}\right) \cdot 
        \left(\frac{i^{|\alpha_n'-\alpha_{n-1}'|}}{\sqrt{2}}\cdots \frac{i^{|\alpha_1'-\alpha_0'|}}{\sqrt{2}}\right).
\end{equation}
By changing the name of the site ``$0$" to ``$-1$", 
RHS of Eq.~(\ref{umetani}) can be reexpressed by 
\[ D(\omega,\omega')= \langle P_{\alpha_n}\cdots P_{\alpha_1}\bs{e}_{\alpha_0}, P_{\alpha_n'}\cdots P_{\alpha_1'}\bs{e}_{\alpha_0'}\rangle \]
with $a=d=1/\sqrt{2}$, $b=c=i/\sqrt{2}$ in Eq. (\ref{qc}) and $\alpha_0'=\alpha_0=-1$. 
This is a special case of $D_{A_0^{(n)}}$. 

\item
$A_P^{(n)}=\{(\xi,\eta)\in A_0^{(n)}: \sum_{j=1}^n\xi_j=\sum_{j=1}^n\eta_j \} $ 

Let $X_n^{(\varphi_0)}$ be a QW at time $n$ starting from the origin with initial state $\bs{\varphi}_0$. Then we define the QW as follows: 
\[ P(X_n^{(\varphi_0)}=x)=\sum_{\xi,\eta}D_{A_P^{(n,x)}}(\xi,\eta), \]
where $A_P^{(n,x)}=\{(\xi,\eta)\in A_P^{(n)}: \sum_{j=1}^n\xi_j=\sum_{j=1}^n\eta_j=x \}$. 
This is consistent with the original definition of the QWs \cite{Gudder1,Meyer}. 
Note that 
\begin{equation}\label{decom}
D_{A_P^{(n)}}
\cong 
\left[
\begin{array}{ccccc} 
D_{A^{(n,-n)}_P} & 0 & & & \\ 
0 & D_{A^{(n,-n+2)}_P} & 0 & & \\ 
 & 0 & \ddots & \ddots & \\ 
 & & \ddots & D_{A^{(n,n-2)}_P} & 0 \\ 
 & & & 0 & D_{A^{(n,n)}_P}
\end{array} 
\right], 
\end{equation}
where $M_1 \cong M_2$ means that there exists a permutation matrix $P$ such that $M_1=PM_2P^{\dagger}$. 
\item
$A_1^{(n)}=\{(\xi,\eta)\in \Omega_n\times\Omega_n: \xi=\eta \} $ 

There are no correlations between two paths except oneself in $A_1^{(n)}$. 
It corresponds to a classical case. 
\end{enumerate}

\noindent In the following, we give examples for the Hadamard coin case and $n=3$ with $\bs{\varphi}_0={}^T[1/\sqrt{2},i/\sqrt{2}]$: 
the order of $2^3$-path in this matrix is $((-1,-1,-1)$, $(-1,-1,1)$, $(-1,1,-1)$, $(1,-1,-1)$, $(-1,1,1)$, $(1,-1,1)$, $(1,1,-1)$, $(1,1,1))$.
\[D_{A_0^{(3)}}=\frac{1}{2^3}
\left[
\begin{array}{c|ccc|ccc|c}
 1 & i & 1 & 0 & -i & 0 & 0 & 0 \\ \hline
 -i & 1 & -i & 0 & -1 & 0 & 0 & 0 \\
 1 & i & 1 & 0 & -i & 0 & 0 & 0 \\
 0 & 0 & 0 & 1 & 0 & i & -1 & i \\ \hline
 i & -1 & i & 0 & 1 & 0 & 0 & 0 \\
 0 & 0 & 0 & -i & 0 & 1 & i & 1 \\
 0 & 0 & 0 & -1 & 0 & -i & 1 & -i \\ \hline
 0 & 0 & 0 & -i & 0 & 1 & i & 1
\end{array}
\right],\;\;\;
D_{A_P^{(3)}}=\frac{1}{2^3}
\left[
\begin{array}{c|ccc|ccc|c}
1 & 0 & 0 & 0 & 0 & 0 & 0 & 0 \\ \hline 
0 & 1 & -i & 0 & 0 & 0 & 0 & 0 \\
0 & i & 1 & 0 & 0 & 0 & 0 & 0 \\
0 & 0 & 0 & 1 & 0 & 0 & 0 & 0  \\ \hline
0 & 0 & 0 & 0 & 1 & 0 & 0 & 0 \\
0 & 0 & 0 & 0 & 0 & 1 & i & 0 \\
0 & 0 & 0 & 0 & 0 & -i & 1 & 0  \\ \hline
0 & 0 & 0 & 0 & 0 & 0 & 0 & 1 
\end{array}
\right],\]
\[D_{A_1^{(3)}}=\frac{1}{2^3}
\left[
\begin{array}{c|ccc|ccc|c}
1 & 0 & 0 & 0 & 0 & 0 & 0 & 0 \\ \hline 
0 & 1 & 0 & 0 & 0 & 0 & 0 & 0 \\
0 & 0 & 1 & 0 & 0 & 0 & 0 & 0 \\
0 & 0 & 0 & 1 & 0 & 0 & 0 & 0  \\ \hline
0 & 0 & 0 & 0 & 1 & 0 & 0 & 0 \\
0 & 0 & 0 & 0 & 0 & 1 & 0 & 0 \\
0 & 0 & 0 & 0 & 0 & 0 & 1 & 0  \\ \hline
0 & 0 & 0 & 0 & 0 & 0 & 0 & 1 
\end{array}
\right].\]
Since $A\subseteq B$ implies $A\prec B$, 
as a subset is smaller, its decoherence matrix becomes scarce. \\ 

\noindent Define $S_A=-\sum_{\lambda\in \mathrm{spec}(D_A)}\lambda \log_2\lambda$, where $\mathrm{spec}(M)$ is the set of eigenvalues of $M$. 
Note that $0\log_2 0=1$ for convention. 
Now we consider the von Neumann entropy $S_{A_0^{(n)}}$, $S_{A_P^{(n)}}$, $S_{A_1^{(n)}}$ where each quantum coin is given by Eq. (\ref{qc}), respectively. 
The following main theorem gives the first and second leading orders of the von Neumann entropy in the limit of $n\to\infty$. 
\begin{theorem}
Let the initial state be $\bs{\varphi}_0={}^T[\alpha,\beta]$ with $|\alpha|^2+|\beta|^2=1$. 
Put $p=|a|^2=|d|^2$, $q=1-p$ and $p_0=|c\alpha+d\beta|$, $q_0=1-p_0$. Then we have 
\begin{enumerate}
\renewcommand{\labelenumi}{\theenumi.}
\item $A_0^{(n)}$ case. 
\begin{align}
&\lim_{n\to \infty}S_{A_0^{(n)}} = 1, \\
&\lim_{n\to \infty}\frac{S_{A_0^{(n)}}-1}{(p-q)^{2n}} = -\left(\frac{p_0-q_0}{p-q}\right)^2\log_2e .
\end{align}
\item $A_P^{(n)}$ case.
\begin{align} 
&\lim_{n\to\infty}\frac{S_{A_P^{(n)}}}{\log_2\sqrt{n}} = 1, \\ 
&\lim_{n\to\infty}\left( \frac{S_{A_P^{(n)}}}{\log_2\sqrt{n}}-1 \right)\log_2 \sqrt{n} =1+\log_2\sqrt{\frac{p}{q}}+\log_2\sqrt{2\pi e}. \label{toshi}
\end{align}
\item $A_1$ case.
\begin{align} 
&\lim_{n\to\infty}\frac{S_{A_1^{(n)}}}{|p\log_2p+q\log_2q|n} = 1, \\ 
&\lim_{n\to\infty}\left( \frac{S_{A_1^{(n)}}}{|p\log_2p+q\log_2q|n}-1 \right)n =\frac{p_0\log_2p_0+q_0\log_2q_0}{p\log_2p+q\log_2q}. \label{tomoya}
\end{align}
\end{enumerate}
\end{theorem}
The first case in the above theorem yields that the von Neumann entropy goes to $1$ with exponentially fast as $n\to \infty$. 
The second and third cases imply that the von Neumann entropy increases with the orders $\log n$ and $n$, respectively as $n$ goes to infinity. 
Moreover the second order of the second and third cases can be described explicitly in Eqs. (\ref{toshi}) and (\ref{tomoya}). 

\section{Proof of theorem}
At first, we introduce the following correlated random walk (RW) which will be useful for the proofs of all the parts 1, 2, and 3. 
For the initial step $n=0$, a particle is located in the origin, and moves to left and right with probabilities $q_0$ and $p_0$, respectively. 
Then for $n\geq 1$, 
the probabilities that a particle chooses the same and other directions of the previous step are $p$ and $q$, respectively at each time step. 
Thus the pair of parameters $(p_0,p)$, which gives the initial condition $\bs{\phi}={}^T[1-p_0, p_0]$ and the stochastic matrix $M$, determines this correlated RW, 
where 
\begin{equation}\label{hikaru} 
M=\begin{bmatrix} p & q \\ q & p \end{bmatrix} 
\end{equation}
with $p+q=1$. We call this walk $(p_0,p)$-correlated RW. 
\begin{lemma}\label{acco1}
Put $\widetilde{P}_j=\bs{e}_j\bs{e}_j^\dagger M$ and $\bs{\phi}_j=\bs{e}_j\bs{e}_j^\dagger\bs{\phi}$ $(j\in \{-1,1\})$. 
Then we have for any $\xi\in \Omega_n$ and the initial state $\bs{\varphi}_0={}^T[\alpha,\beta]$, in the case of $p=|a|^2$, $p_0=|c\alpha+d\beta|^2$, 
\begin{equation}
||\bs{w}^{(\varphi_0)}(\xi)||^2=\langle \bs{1}, \widetilde{P}_{\xi_n}\cdots \widetilde{P}_{\xi_2}\bs{\phi}_{\xi_1}\rangle, 
\end{equation}
where $\bs{1}$ is the all 1 vector. 
\end{lemma}
{\it Proof.}
Recall the definition of weight of path $\xi=(\xi_n,\dots,\xi_1)\in \Omega_n$ and $P_j$. 
Then we obtain 
\begin{align}\label{pipi1}
||\weight (\xi)||^2 
	&= ||P_{\xi_n}\cdots P_{\xi_1}\bs{\varphi}_0||^2 \notag \\
        &= |\langle \bs{e}_{\xi_n}, U \bs{e}_{\xi_{n-1}}\rangle|^2 
        	\cdots |\langle \bs{e}_{\xi_2}, U\bs{e}_{\xi_1}\rangle|^2|\langle \bs{e}_{\xi_1}, U \bs{\varphi}_0\rangle|^2 .
\end{align}
The unitarity of the quantum coin $U$ provides that 
\begin{align} 
|\langle \bs{e}_{l}, U\bs{e}_{m}\rangle|^2 
	&= \delta_{l,m}p+(1-\delta_{l,m})q=(M)_{l,m}, \label{pipi2}\\
|\langle \bs{e}_{l}, U \bs{\varphi}_0\rangle|^2 
	&=
        \begin{cases}
        p_0 & \text{: $l=1$,} \\ q_0 & \text{: $l=-1$,}
        \end{cases} \label{pipi3}
\end{align}
where $p=|a|^2=|d|^2$, $q=|b|^2=|c|^2$ and $p_0=|c\alpha+d\beta|^2$, $q_0=|a\alpha+b\beta|^2$ with $p+q=p_0+q_0=1$. 
Combining Eq. (\ref{pipi1}) with Eqs. (\ref{pipi2}) and (\ref{pipi3}), 
we can interpret $||\weight (\xi)||^2$ as the probability that $(p_0,p)$-correlated random walker with parameters $p=|a|^2$ and $p_0=|c\alpha+d\beta|^2$ 
walks along with the path $\xi$. Then we arrive at the desired conclusion. 
\begin{flushright} $\square$ \end{flushright} 
We also use the following lemma obtained by a standard argument of the matrix analysis. 
\begin{lemma}\label{acco2}
For any $m\times n$ matrix $M$, the two matrices  
$MM^\dagger$ and $M^\dagger M$ have the same non-zero eigenvalues except $0$. 
\end{lemma}
\subsection{Proof of part 1}
Define $\{ \xi^{(k)} \}_{k=1}^{m}$ and $\{ \eta^{(k)} \}_{k=1}^{m}$ as the set of all the $n$-truncated paths 
whose finial directions are the left and the right, respectively, 
that is, $\xi^{(k)}(n)=-1$ and $\eta^{(k)}(n)=1$ for any $k$. Here $m=|\Omega_n|/2=2^{n-1}$. 
Let $T_{A_0^{(n)}}$ be a $2\times 2^n$ matrix defined by
\[ T_{A_0^{(n)}}=\left[\bs{w}^{(\varphi_0)}(\xi^{(1)}),\dots,\bs{w}^{(\varphi_0)}(\xi^{(m)}),
	\bs{w}^{(\varphi_0)}(\eta^{(1)}),\dots,\bs{w}^{(\varphi_0)}(\eta^{(m)})\right] \]
Then we get 
\begin{equation}\label{tokuzo1} 
D_{A_0^{(n)}}\cong T_{A_0^{(n)}}^\dagger T_{A_0^{(n)}},
\end{equation}
where ``$A \cong B$" means that there exists a permutation operator $P$ on $\Omega_n$ such that $B=P^\dagger A P$. 
Noting Eq. (\ref{pero}), 
\begin{align}\label{tokuzo2}
T_{A_0^{(n)}}T_{A_0^{(n)}}^\dagger 
	&= \sum_{k=1}^{m}\bs{w}^{(\varphi_0)}(\xi^{(k)})\bs{w}^{(\varphi_0)}(\xi^{(k)})^\dagger
		+\sum_{k=1}^{m}\bs{w}^{(\varphi_0)}(\eta^{(k)})\bs{w}^{(\varphi_0)}(\eta^{(k)})^\dagger \notag \\
        &=
        \begin{bmatrix}
        \rho_{L}^{(n)} & 0 \\
	0 & \rho_{R}^{(n)}
	\end{bmatrix}
\end{align}
where $\rho_L^{(n)}=\sum_{k=1}^{m}||\weight(\xi^{(k)})||^2$, $\rho_R^{(n)}=\sum_{k=1}^{m}||\weight(\eta^{(k)})||^2$. 
From Lemma \ref{acco1}, 
$\rho_{L}^{(n)}$ (resp. $\rho_R^{(n)}$) is the probability that a correlated random walker arrives at the final position 
from the left (resp. right), respectively. 
Now we compute the probabilities $\rho_L^{(n)}$ and $\rho_R^{(n)}$. 
Since $\widetilde{P}_{1}+\widetilde{P}_{-1}=M$, we see 
\begin{align*}
\rho_L^{(n)} &= \sum_{\xi_1,\dots,\xi_{n-1}\in \{\pm 1\}} 
		\left \langle \bs{e}_{-1}, 
			\widetilde{P}_{\xi_n}\cdots \widetilde{P}_{\xi_2}\bs{\phi}_{\xi_1}  \right\rangle
	   = \left \langle \bs{e}_{-1}, M^{n-1}\bs{\phi} \right\rangle, \\
\rho_R^{(n)} &= \left \langle \bs{e}_{1}, M^{n-1}\bs{\phi} \right\rangle.
\end{align*}
The eigenvalues and eigenvectors of $M$ are $\{1,p-q\}$ and corresponding eigenvectors are ${}^T[1/\sqrt{2},1/\sqrt{2}]$ 
and ${}^T[1/\sqrt{2},-1/\sqrt{2}]$, respectively. 
Therefore we obtain 
\begin{equation}\label{hiroshi}
\rho_L^{(n)}=\frac{1}{2}\left\{ 1+(p-q)^{n-1}(p_0-q_0) \right\},\;\;
\rho_R^{(n)}=\frac{1}{2}\left\{ 1-(p-q)^{n-1}(p_0-q_0) \right\}.
\end{equation}
Substituting Eq. (\ref{hiroshi}) into Eq. (\ref{tokuzo2}), 
Lemma \ref{acco2} and Eq. (\ref{tokuzo1}) give the eigenvalues of $D_{A_0^{(n)}}$ as follows: 
\begin{equation}
\mathrm{spec}(D_{A_0^{(n)}})=\left\{ \rho_L^{(n)}, \rho_R^{(n)},\overbrace{0,\dots,0}^{2^n-2} \right\}.
\end{equation}
The von Neumann entropy of $D_{A_0^{(n)}}$ can be described by
\begin{align}
 S_{A_0^{(n)}} &= -\left(\rho_L^{(n)}\log_2 \rho_L^{(n)}+\rho_R^{(n)}\log_2 \rho_R^{(n)}\right). \label{haruki} 
\end{align}
Noting $\rho_J^{(n)}$ ($J\in\{L,R\}$) converges to $1$ ($n\to\infty$) exponentially fast with the base $(p-q)$ by Eq. (\ref{hiroshi}), 
then we have $S_{A_0^{(n)}} \to 1\;\; (n\to\infty)$. 

Moreover from the Taylor expansions around $1$ of the logarithm of Eq. (\ref{hiroshi}), we give for large $n$, 
\begin{equation}\label{seigo}
1+\log_2\rho_L^{(n)} \sim (p_0-q_0)(p-q)^{n-1}\log_2e,\;\mathrm{and}\; 1+\log_2\rho_R^{(n)}\sim - (p_0-q_0)(p-q)^{n-1}\log_2e, 
\end{equation}
where $a_n\sim b_n$ means $\lim_{n\to \infty} |a_n/b_n|=1$. 
By substituting Eq. (\ref{seigo}) into RHS of Eq. (\ref{haruki}), we obtain $1-S_{A_0^{(n)}}\sim (p_0-q_0)^2(p-q)^{2(n-1)}\log_2e$ 
which completes the proof of part 1. 
\begin{flushright} $\square$ \end{flushright}
\subsection{Proof of part 2}
Define $T_{A_P^{(n,j)}}$ by 
\[ T_{A_P^{(n,j)}}=\begin{bmatrix} \weight(\xi^{(1)}),\dots,\weight(\xi^{(l)}),\weight(\eta^{(1)}),\dots,\weight(\eta^{(m)}) \end{bmatrix}, \]
where $-n\leq j\leq n$ and 
\begin{align*}
\{\xi^{(k)}\}_{k=1}^{l} &= \left\{\xi\in \Omega_n: \sum_{i=1}^{n}\xi_i=j\;\mathrm{and}\; \xi_n=-1  \right\}, \\
\{\eta^{(k)}\}_{k=1}^{m} &= \left\{\eta\in \Omega_n: \sum_{i=1}^{n}\eta_j=j\;\mathrm{and}\; \eta_n=1  \right\}.
\end{align*}
Then we have $D_{A_P^{(n,j)}}\cong T_{A_P^{(n,j)}}T_{A_P^{(n,j)}}^\dagger$. 
By a similar fashion of the proof of part 1, 
we obtain 
\begin{equation}
\mathrm{spec}\left(D_{A_P^{(n,j)}}\right)=\left\{ p_L^{(n)} (j),p_R^{(n)} (j),\overbrace{0,\dots,0}^{\binom{n}{(n+j)/2}-2}  \right\},
\end{equation}
where $p_L^{(n)}(j)$ (resp. $p_R^{(n)}(j)$) is the probability that a $(p_0,p)$-correlated random walker with parameters $p=|a|^2$ and $p_0=|c\alpha+d\beta|^2$ 
arrives at position $j$ from the left (resp. right) direction at time $n$, respectively, that is, 
\begin{equation}
p_L^{(n)}(j) = \sum_{\xi: \sum_{k=1}^n\xi_k=j} \langle \bs{e}_{-1}, \widetilde{P}_{\xi_n}\cdots \widetilde{P}_{\xi_2}\bs{\phi}_{\xi_1} \rangle, \;\;
p_R^{(n)}(j) = \sum_{\xi: \sum_{k=1}^n\xi_k=j} \langle \bs{e}_1, \widetilde{P}_{\xi_n}\cdots \widetilde{P}_{\xi_2}\bs{\phi}_{\xi_1} \rangle.
\end{equation} 
From now on, we compute the asymptotic behaviors of $p_L^{(n)}(j)$ and $p_R^{(n)}(j)$ in the limit of $n\to\infty$ by using the Fourier transform. 
Let $\bs{\Psi}^{(n)}(j)={}^T[p_L^{(n)}(j),p_R^{(n)}(j)]$ for $n\geq 1$. Define the spatial Fourier transform of $\bs{\Psi}^{(n)}(j)$ by 
$\widehat{\bs{\Psi}}^{(n)}(\xi)=\sum_{j\in \mathbb{Z}}\bs{\Psi}^{(n)}(j)e^{i\xi j}$ for $\xi\in [0,2\pi)$. 
From the definition of $(p_0,p)$-correlated RW, 
\begin{align}
\bs{\Psi}^{(1)}(j) &= \delta_{\{j=1\}}\; p_0 \bs{e}_1+\delta_{\{j=-1\}}\;q_0 \bs{e}_{-1}, \\
\bs{\Psi}^{(n)}(j) &= \widetilde{P}_1\bs{\Psi}^{(n-1)}(j-1)+\widetilde{P}_{-1}\bs{\Psi}^{(n-1)}(j+1)\;\;(n\geq 2).
\end{align}
We have 
\begin{equation}\label{paul}
\widehat{\bs{\Psi}}^{(n)}(\xi)
	= \widehat{M}^{n-1}(\xi) \begin{bmatrix} q_0e^{-i\xi} \\ p_0e^{i\xi} \end{bmatrix},
\end{equation}
where 
\[ \widehat{M}(\xi)=\begin{bmatrix}e^{-i\xi} & 0 \\ 0 & e^{i\xi}\end{bmatrix}M. \]
The eigenvalue $\lambda_\xi^{(\pm)}$ and its eigenvector $\bs{v}_\xi^{(\pm)}$ of $\widehat{M}(\xi)$ are 
\[ \lambda_\xi^{(\pm)}=p\cos\xi \pm \sqrt{q^2-p^2\sin \xi},\;\; 
\bs{v}_\xi^{(\pm)}=\frac{1}{\Lambda_\xi^{\pm}}\begin{bmatrix} e^{i\xi}q \\ \lambda_\xi^{(\pm)}-e^{i\xi}p \end{bmatrix}, \]
where $\Lambda_\xi^{(\pm)}$ is the normalized constant. 
Replacing $\xi$ to $\xi/\sqrt{n}$ provides the following asymptotics of the eigensystem for large $n$ as follows.  
\begin{equation}\label{kaoru} 
\lambda_{\xi/\sqrt{n}}^{(+)} \sim 1-\frac{p/q}{2n}\xi^2,\;\;
\lambda_{\xi/\sqrt{n}}^{(-)} \sim (p-q)\left(1+\frac{p/q}{2n}\xi^2\right). 
\end{equation} 
Substituting Eq. (\ref{kaoru}) into Eq. (\ref{paul}) gives for large $n$, 
\begin{equation}
\widehat{\bs{\Psi}}^{(n)}(\xi/\sqrt{n})\sim \frac{1}{2}e^{-\frac{p/q}{2}\xi^2}\bs{1}. 
\end{equation}
Therefore we have 
\begin{equation}\label{eiri}
\lim_{n\to\infty}\sum_{j\leq \sqrt{n}x} p^{(n)}_J(j)
	=\frac{1}{2} \int_{-\infty}^{x} \frac{e^{-\frac{y^2}{2p/q}}}{\sqrt{2\pi p/q}}dy, \;\;(J\in \{L,R\}).
\end{equation}
Equation (\ref{eiri}) implies that 
\begin{equation}\label{mitsuru} 
p^{(n)}_J(j)\sim \frac{1}{2}\frac{e^{-\frac{x^2}{2p/q}}}{\sqrt{2\pi n p/q}},\;\;(x=j/\sqrt{n}). 
\end{equation}
Noting Eq. (\ref{decom}), then it is obtained that 
\begin{align*}
-S_{A_p^{(n)}} &= \sum_{j=-n}^n p^{(n)}_L(j) \log_2 p^{(n)}_L(j)+\sum_{j=-n}^n p^{(n)}_R(j) \log_2 p^{(n)}_R(j)  \notag \\
	       &\sim 2\times \int_{-\infty}^\infty \frac{1}{2}\frac{e^{-\frac{x^2}{2p/q}}}{\sqrt{2\pi p/q}}
               		\log_2 \left(\frac{1}{2}\frac{e^{-\frac{x^2}{2p/q}}}{\sqrt{2\pi n p/q}}\right) dx \notag \\
               &\;\;\;\;\;\;\;\;\;= -\left(1+\log_2 \sqrt{n} +\log_2\sqrt{\frac{p}{q}}+\log_2 \sqrt{2\pi e}\right).         
\end{align*} 
Then we have the desired conclusion. 
\begin{flushright} $\square$ \end{flushright}
\begin{corollary}\label{kayoko}
Put $Y_n$ be the correlated RW at time $n$. Since $P(Y_n=j)=p^{(n)}_L(j)+p^{(n)}_R(j)$, 
Eq. (\ref{mitsuru}) gives the asymptotics of the Shannon entropy of the correlated RW, $H_n^{(RW)}=-\sum_{j=-n}^nP(Y_n=j)\log P(Y_n=j)$, 
in the limit of $n\to \infty$ as follows: 
\[ H_n^{(RW)}\sim \log_2 \sqrt{n} +\log_2\sqrt{\frac{p}{q}}+\log_2 \sqrt{2\pi e}.  \]
\end{corollary}
\subsection{Proof of part 3}
It should be noticed that since $D_{A_1^{(n)}}$ is the diagonal matrix with $D_{A_1^{(n)}}(\xi,\xi)=||\weight (\xi)||^2$ ($\xi \in \Omega_n$), 
\begin{equation}\label{ikumi} 
\mathrm{spec}(D_{A_1^{(n)}})=\{||\weight (\xi)||^2: \xi \in \Omega_n\}. 
\end{equation}
We should recall the definition of the correlated RW in this paper: 
At the first step, a walker moves to the left and right directions with probabilities $q_0$ and $p_0$, respectively. 
Since then, if a choice of the directions is changed, then its associated probability is $q$, otherwise $p$. 
Therefore the absolute value of the weight of path $\xi$ is determined by the first choice of direction and the number of changes of the directions. 
Noting that the maximal number of the changes is $n-1$, 
if the number of changes of directions is $j$, then the number of such paths is $\binom{n-1}{j}$ and 
\begin{equation}\label{wataru}
||\weight (\xi)||^2=
	\begin{cases}
        p_0p^{n-1-j}q^j & \text{: the first choice of directions is right, }\\
        q_0p^{n-1-j}q^j & \text{: the first choice of directions is left. }
        \end{cases}
\end{equation}
Substituting Eq. (\ref{wataru}) into Eq. (\ref{ikumi}), we obtain 
\begin{equation}\label{d1} 
\mathrm{spec}(D_{A_1^{(n)}})=\left\{ p_0p^{n-1-j}q^j, q_0p^{n-1-j}q^j \;\;\mathrm{with\;each\;multiplicity\;} \binom{n-1}{j}: 0\leq j\leq n-1\right\}.  
\end{equation}
Combining Eq. (\ref{d1}) with some properties of the binomial distribution $B(n-1,p)$, we get 
\[ S_{A_1^{(n)}}=(n-1)\left|p\log_2 p+q\log_2 q\right|+\left|p_0\log_2 p_0+q_0\log_2 q_0\right|. \]
\begin{flushright} $\square$ \end{flushright}
\section{Summary and discussion}
We computed the von Neumann entropy of the decoherence matrix studied by Gudder and Sorkin~\cite{GS}, 
which is restricted to three subsets $A_0^{(n)}$, $A_P^{(n)}$, and $A_1^{(n)}$, respectively, in the $n$-truncated path space. 
We found that all the eigenvalues of each decoherence matrix are expressed by the probability that the $(p_0,p)$-correlated random walker 
chooses the path corresponding to each subset. 
We showed that each von Neumann entropy of the decoherence matrix restricted by 
$A_1^{(n)}\prec A_P^{(n)}\prec A_0^{(n)}$ 
is $\sim 1$, $\sim \log n$, and $\sim n$, respectively, asymptotically for large $n$. 
The subset $A_P^{(n)}$ corresponds to the QW, while $A_1^{(n)}$ corresponds to the correlated RW on $\mathbb{Z}$. 
In Ref.\cite{IKM}, the Shannon entropy of the QW, $H_n^{(QW)}=-\sum_j P(X_n^{(\varphi_0)}=j)\log P(X_n^{(\varphi_0)}=j)$, 
is obtained asymptotically with $H_n^{(QW)}\sim \log n$. 
On the other hand, from Corollary \ref{kayoko}, 
its corresponding Shannon entropy of the correlated RW is $H_n^{(RW)}\sim \log \sqrt{n}$. 
We summarize these results in Table \ref{sum}. 
\begin{table}
\begin{center}
\begin{tabular}{|c|c|c|}
\hline
            & Correlated RW & QW \\ \hline
    Shannon entropy & $\sim \log \sqrt{n}$  \;(Cor.\ref{kayoko}) & $\sim \log n$  \;(Ref.~\cite{IKM}) \\ \hline
von Neumann entropy & $\sim -(p\log p+q\log q)n$  \;(Thm.1 (3)) & $\sim \log \sqrt{n}$  \;(Thm.1 (2)) \\ \hline
\end{tabular}
\end{center}
\caption{Summary of the asymptotic behaviors of the Shannon and von Neumann entropy of the QW and the correlated RW}
\label{sum}
\end{table}

From our results, we conjecture that if $A\subseteq A'$, then $S_{A}\geq S_{A'}$.
Indeed, if we take a subset $B$ with $A_1^{(n)}\subseteq B\subseteq A_P^{(n)}$ such that 
\begin{equation}\label{jim} B=\{(\xi,\eta)\in A_P^{(n)}: \xi_1=\eta_1  \},  \end{equation}
then by using a similar fashion of the proof of part 2 in Theorem 1, we get 
\[ \lim_{n\to\infty}\left(\frac{S_B}{\log_2 \sqrt{n}}-1\right)\log_2 \sqrt{n} 
	= 1+\log_2\sqrt{\frac{p}{q}}+\log_2\sqrt{2\pi e}+\left|p_0\log_2 p_0+q_0\log_2 q_0\right|. \] 
Comparing with Eq.~(\ref{toshi}), due to the extra term $|p_0\log_2 p_0+q_0\log_2 q_0|$ of $S_B$, we actually see that
$S_B\geq S_{A_P^{(n)}}$ for large $n$. 
The subset $B$ defined by Eq.(\ref{jim}) gives only an increase of the constant value with respect to the size $n$ in the entropy. 
If we give a smaller subset with $B'\subset B$, 
then when we can see the change of the leading order of the entropy from $\log _2 \sqrt{n}$ to $n$? 
To consider the von Neumann entropy of a decoherence matrix restricted to such a subset $B'$ with 
$A_1^{(n)}{\subseteq} B'{\subseteq} A_0^{(n)}$ is one of the interesting future works.

\par
\
\par\noindent
\noindent
{\bf Acknowledgments.}
NK acknowledges financial support of the Grant-in-Aid for Scientific Research (C) of Japan Society for the
Promotion of Science (Grant No. 21540118).
\par
\
\par

\begin{small}
\bibliographystyle{jplain}

\end{small}

\noindent\\
\noindent\\
\noindent\\



\end{document}